\documentclass[11pt]{article}
\pdfoutput=1

\usepackage{jheppub}
\usepackage{bigints}
\usepackage{rotating}

\usepackage{tabularx}

\usepackage{tikz}
\usepackage{fancybox}
 
\usepackage{tikz-cd}
\usepackage{ae}
\usepackage{url}

\usepackage{tabularx}

\usepackage{graphicx}
\usepackage{dsfont}
\usepackage{setspace}
\usepackage{amsfonts,amsmath,amsthm,amssymb,amsbsy}
\usepackage{longtable}
\usepackage{array}
\usepackage{color}
\usepackage{needspace}
\usepackage[numbers]{natbib}
\usepackage{colortbl}

\usepackage{enumitem}

\usepackage{fancyhdr}

\usepackage{tabularx}

\usepackage{graphicx}
\usepackage{dsfont}
\usepackage{setspace}
\usepackage{amsfonts,amsmath,amsthm,amssymb,amsbsy}
\usepackage{longtable}
\usepackage[latin1]{inputenc}
\usepackage{array}
\usepackage{color}
\usepackage{needspace}
\usepackage[numbers]{natbib}
\usepackage{colortbl}

\usepackage{xcolor}
\definecolor{colourO}{HTML}{4FEEBB}
\definecolor{colourL}{HTML}{5DB197}
\definecolor{colourG}{HTML}{3E7CEF}
\definecolor{colourK}{HTML}{3535E3}

\usepackage{enumitem}

\usepackage{fancyhdr}

  \def\david#1 {\{\color{cyan} DD:#1 \color{black}\}}

\title{Momentum Amplituhedron meets Kinematic Associahedron\footnote{despite Coronavirus}}

\author[1]{David Damgaard,}\emailAdd{d.damgaard@lmu.de}
\author[1,2]{Livia Ferro,}\emailAdd{livia.ferro@lmu.de}
\author[2]{Tomasz \L ukowski,}\emailAdd{t.lukowski@herts.ac.uk}
\author[2]{and Robert Moerman}\emailAdd{r.moerman@herts.ac.uk}

\affiliation[1]{Arnold--Sommerfeld--Center for Theoretical Physics,\\ Ludwig--Maximilians--Universit\"at, \\ Theresienstra\ss e 37, 80333 M\"unchen, Germany }
\affiliation[2]{Department of Physics, Astronomy and Mathematics, \\ University of Hertfordshire, \\  Hatfield, Hertfordshire, AL10 9AB, United Kingdom}

\abstract{In this paper we study a relation between two positive geometries: the momentum amplituhedron, relevant for tree-level scattering amplitudes in $\mathcal{N}=4$ super Yang-Mills theory, and the kinematic associahedron, encoding tree-level amplitudes in bi-adjoint scalar $\phi^3$ theory. We study the implications of restricting the latter to four spacetime dimensions and give a direct link between its canonical form and the canonical form for the momentum amplituhedron. After removing the little group scaling dependence of the gauge theory, we find that we can compare the resulting reduced forms with the pull-back of the associahedron form. In particular, the associahedron form is the sum over all helicity sectors of the reduced momentum amplituhedron forms. This relation highlights the common singularity structure of the respective amplitudes; in particular, the factorization channels, corresponding to vanishing planar Mandelstam variables, are the same. Additionally, we also find a relation between these canonical forms directly on the kinematic space of the scalar theory when reduced to four spacetime dimensions by Gram determinant constraints. As a by-product of our work we provide a detailed analysis of the kinematic spaces relevant for the four-dimensional gauge and scalar theories, and provide direct links between them.  }
\begin{document}
\begin{flushright}
{\small LMU-ASC 41/20}
\end{flushright}
\maketitle



\section{Introduction}

Scattering amplitudes, being the primary building blocks of physical observables, are centrally important in high-energy physics research. Their computation in perturbative quantum field theory is complicated, with difficulties stemming from redundancies which are introduced in order to make locality and unitarity manifest. However, in recent years there has been tremendous development in the techniques available for studying them. These techniques have significantly improved computational efficiency as well as pushed our conceptual understanding of high-energy physics and several related disciplines. One strand of this development, proposed by Cachazo, He and Yuan (CHY), led to a new formulation for the tree-level scattering of massless particles in arbitrary dimensions where amplitudes can be written as integrals over the moduli space of Riemann spheres \cite{Cachazo:2013hca,Cachazo:2013iea}. When restricted to four dimensions, this approach resulted in new twistor-string-inspired formulae for tree-level amplitudes for $\mathcal{N}=8$ supergravity \cite{Geyer:2014fka}, planar $\mathcal{N}=4$ super Yang-Mills theory (sYM) \cite{Geyer:2014fka} and four-dimensional bi-adjoint scalar $\phi^3$ theory \cite{Cachazo:2016sdc}. The latter two theories also feature prominently in another, equally novel approach, where scattering amplitudes are encoded geometrically.
The paradigm-shift which underpins this geometric method is that rather than thinking of scattering amplitudes as functions, they are thought of as differential forms  
on the \emph{kinematic space}, the space of physical kinematic variables. 
The setting where these ideas were first employed was in the study of 
scattering amplitudes in planar $\mathcal{N}=4$ sYM in momentum twistor space. 
Here the kinematic space for $n$ particles is the space of momentum twistors $z_i$ with $i=1,\ldots,n$ and the differential form is found from a scattering amplitude by substituting the Grassmann-odd variables $\eta_i$, related to the $R$-symmetry of the superspace, with differentials of $z_i$:  $\eta_i \rightarrow dz_i$. The geometry associated with this differential form is the amplituhedron \cite{Arkani-Hamed:2013jha}. It is an example of a more general class of objects, collectively called {\it positive geometries} \cite{Arkani-Hamed:2017tmz} --- real, oriented, closed geometries with boundaries of all co-dimension equipped with canonical differential forms, which have  logarithmic singularities along all boundaries. 

Our focus in this paper will be on two particular positive geometries: 
the {\it momentum amplituhedron} \cite{Damgaard:2019ztj}, whose logarithmic differential form gives tree-level amplitudes for $\mathcal{N}=4$ sYM in spinor helicity space; and the {\it kinematic associahedron} \cite{Arkani-Hamed:2017mur}, from which one can extract tree-level amplitudes for bi-adjoint $\phi^3$ theory. Although amplitudes in both these theories can be written as integrals over moduli spaces in the CHY formalism, a direct link has yet to be established between them from the point of view of the geometries. In this paper we fill this gap and present an explicit relation between canonical forms for momentum amplituhedra and those for kinematic associahedra and therefore give a direct link between scattering amplitudes in $\mathcal{N}=4$ sYM and bi-adjoint $\phi^3$ theory.

In order to make this link possible, one needs to study tree-level $\mathcal{N}=4$ sYM amplitudes in non-chiral superspace. Here a tree-level amplitude in the  helicity-$k$ sector can be written as a differential form of degree $(2(n-k),2k)$ in $(d\lambda,d\widetilde\lambda)$ through the replacement $(\eta,\widetilde\eta) \to (d\lambda\,, d\widetilde\lambda)$ \cite{He:2018okq}. This differential form, after we factorize (half of) the supermomentum conservation, agrees with the canonical form $\Omega_{n,k}$ of the momentum amplituhedron $\mathcal{M}_{n,k}^{(\lambda,\tilde\lambda)}$, which has degree $(2n-4)$ \cite{Damgaard:2019ztj}.  Since this degree does not depend on $k$, it is possible to define a form for the full $n$-point superamplitude as the sum of the canonical forms for the $n$-point amplitudes in the different $k$-sectors. In the non-chiral superspace, the superamplitude $\mathbb{A}_n$ is invariant under the $GL(1)^n$ little group scaling and it is natural to describe it using a little group invariant parametrization of the on-shell space containing cross-ratios of spinor helicity variables and Mandelstam variables. When we use this parametrization for the momentum amplituhedron canonical form instead, this defines a unique, little group invariant form of degree $(n-3)$, which we call the {\it reduced momentum amplituhedron form} $\omega_{n,k}$. Roughly speaking, this parametrization leads to the relation
\begin{equation}
\Omega_{n,k}=\mu_n\wedge \omega_{n,k}+\ldots\,,
\end{equation} 
where $\mu_n$ is the canonical top form of degree $n-1$ on the projective space $\mathbb{P}^{n-1}$, depending only on the little group scaling parameters, which can be naturally associated with the action of the little group, and the ellipsis $(\ldots)$ contains terms which are not invariant under little group scaling. On the other hand, the kinematic associahedron is a geometry on the kinematic space parametrized solely by planar Mandelstam variables and its canonical differential form also has degree $(n-3)$. The main result of this paper is that these forms are related to each other\footnote{Indeed, at the level of the degree of the differential forms involved, $(2n-4)=(n-1)+(n-3)$.}! More precisely, when we evaluate the associahedron form on the little group invariant space, then it is the same as the sum of reduced momentum amplituhedron forms summed over all helicity sectors. This relation exposes the common physics of tree-level amplitudes in both theories and reflects the fact that some singularities of $\mathcal{N}=4$ sYM are given by planar tree cubic graphs. Indeed, tree-level color-ordered amplitudes in $\mathcal{N}=4$ sYM have poles where the sum of adjacent momenta, the planar Mandelstam variables, goes on shell $P_{i,j}^2=(p_i+p_{i+1}+\ldots+p_j)^2\rightarrow 0$,
 which agrees with the factorization singularities of the double colour-ordered amplitudes in bi-adjoint $\phi^3$ theory, where poles are given by the vanishing of the propagators $P_{i,j}^2$.

One additional question arising from our analysis is whether it is possible to compare the reduced momentum amplituhedron and associahedron forms directly in the space of planar Mandelstam variables. We can answer this question in the affirmative for four- and five-particle scattering, but it is not possible to make such a comparison for higher numbers of particles. This descends from the fact that in four dimensions any set of at least six on-shell momenta is not independent due to the {\it Gram determinant constraints}. We will however be able to compare the reduced momentum amplituhedron and associahedron forms on the kinematic space of the scalar theory modulo these Gram determinant constraints. In order to make this comparison we will compute their push-forwards on this reduced space and find that the push-forward of the associahedron form is proportional to the sum over all helicities of push-forwards of the reduced momentum amplituhedron forms, with a known proportionality factor.

The paper is structured as follows. We begin by reviewing the definitions of the momentum amplituhedron and the kinematic associahedron in section \ref{sec:definitions}. Both cases  
can be described as the intersection of a top-dimensional ``positive" region -- a region of kinematic space constrained by particular positivity conditions -- and a family of affine subspaces of the kinematic space. In section \ref{sec:maps}, we define various kinematic spaces relevant to our discussion and the maps relating them. We also describe the action that these maps induce on differential forms. Thereafter, we discuss the main result of this paper, i.e.\ how the canonical forms of the momentum amplituhedron and the kinematic associahedron are related to each other.
Then, in section \ref{sec:examples}, we examine these general statements in explicit examples. Finally, in section \ref{sec:ISF}, we discuss a general procedure for constructing the reduced momentum amplituhedron forms. After the conclusions, the appendices collect various technical results and formulae.


\section{Definitions}
\label{sec:definitions}
We begin by reviewing the two positive geometries which will be considered in this paper: the momentum amplituhedron \cite{Damgaard:2019ztj} and the kinematic associahedron \cite{Arkani-Hamed:2017mur}.  Every positive geometry comes equipped with a canonical differential form, whose {\it leading singularities} or residues on zero-dimensional boundaries equal   $\pm 1$. The canonical differential form of the momentum amplituhedron, respectively associahedron, encodes the tree-level scattering amplitudes for $\mathcal{N}=4$ sYM, respectively bi-adjoint $\phi^3$, theory.
In both cases, we provide their definitions in their respective kinematic space in terms of the intersection of a positive region with an affine subspace. For an extensive review on these and other positive geometries, see \cite{Ferro:2020ygk}.


\subsection{Momentum Amplituhedron}

The momentum amplituhedron $\mathcal{M}^{(\lambda,\widetilde\lambda)}_{n,k}$ is the positive geometry associated with tree-level scattering amplitudes in $\mathcal{N}=4$ sYM in spinor helicity space \cite{Damgaard:2019ztj}.
Superamplitudes in $\mathcal{N}=4$ sYM are defined for on-shell chiral superfields $\Phi_i$,  which collect the on-shell supermultiplet into a single object by means of four Grassmann-odd variables $\eta^A_i$, $A=1,\ldots,4$, for each particle $i$.  A generic $n$-particle superamplitude $\mathbb{A}_n = \mathbb{A}_n(\Phi_1, \Phi_2, \ldots,\Phi_n)$ can be expanded in terms of helicity sectors, denoted by $k$, of Grassmann degree $4k$:
\begin{equation}
\mathbb{A}_n = A_{n,2} +  A_{n,3} + \ldots +  A_{n,n-2},\quad n\geq 4\,,
\end{equation}
where $A_{n,2}$ is the maximally-helicity-violating (MHV) amplitude,  $ A_{n,3}$ is the next-to-MHV (NMHV) amplitude and so on, with $A_{n,k}$ the amplitude for the $\text{N}^{k-2}\text{MHV}$ sector. 
Therefore the superamplitudes live in the on-shell chiral superspace $(\lambda^a,\widetilde\lambda^{\dot a}|\eta^A)$, $a,\dot a=1,2$. However, in order to interpret them as differential forms and define the associated positive geometry, we need to rewrite them in    
the non-chiral superspace $(\lambda^a,\eta^r \,|\, \widetilde\lambda^{\dot a},\widetilde\eta^{\dot r})$, $r,\dot r=1,2$, where we perform a Fourier transform for two of the four Grassmann-odd variables. 
 In this way,  via  the replacement
\begin{equation}
\label{lambda_eta}
\eta^a \to d\lambda^a\, ,\qquad\qquad \widetilde\eta^{\dot a}\to d\widetilde\lambda^{\dot a}\,,
\end{equation} 
the tree-level N$^{k-2}$MHV  scattering amplitudes can be written as differential forms of degree $(2(n-k),2k)$ in $(d\lambda,d\widetilde\lambda)$  \cite{He:2018okq}. The geometry whose canonical differential form is this tree-level amplitude form is the momentum amplituhedron.

We can define $\mathcal{M}^{(\lambda,\widetilde\lambda)}_{n,k}$ directly in terms of kinematic data in the spinor helicity space, without any reference to auxiliary spaces.  Let us start by defining the following  $(2n-4)$-dimensional subspace of the kinematic space
\begin{equation}\label{subspace.definition.mom}
\mathcal{V}_{n,k} 
:= 
 \{(\lambda_i^a,\widetilde\lambda_i^{\dot a}):
     \lambda_i^a = \lambda^{*a}_i+y_{\alpha}^a \,\Delta_i^\alpha ,       \widetilde\lambda_i^{\dot a} = \widetilde\lambda^{*\dot{a}}_i+\widetilde y_{\dot{\alpha}}^{\dot{a}} \,\widetilde\Delta_i^{\dot\alpha},\lambda_i^a \widetilde\lambda_i^{\dot a}=0 
\}\,,
\end{equation}
where $(\lambda^*,\widetilde\lambda^*)$ are two fixed two-planes in $n$ dimensions, $\widetilde\Delta$ is a fixed $k$-plane and $\Delta$ is an $(n-k)$-dimensional fixed plane in $n$ dimensions. Moreover, we assume that when we assemble these subspaces as in 
\begin{equation}\label{big.lambda}
\Lambda_i^A = 
\begin{pmatrix} 
\lambda_i^{a *} \\
\Delta^\alpha_i
\end{pmatrix} ,
\qquad \widetilde\Lambda_i^{\dot A} = 
\begin{pmatrix} 
\widetilde\lambda_i^{\dot a *} \\
\widetilde\Delta^{\dot\alpha}_i
\end{pmatrix}  \, ,
\end{equation}
 $\widetilde\Lambda$ is a positive matrix and $\Lambda$ is a twisted positive matrix; see \cite{Lukowski:2020dpn} for a precise definition of the latter.  We also define a winding space $\mathcal{W}_{n,k}$ as the subset of kinematic space 
satisfying the conditions \cite{He:2018okq}
\begin{align}
\mathcal{W}_{n,k}:=&\{(\lambda_i^a,\widetilde\lambda_i^{\dot a}):s_{i,i+1,\ldots,i+j}>0\,, \nonumber \\&
\mbox{the sequence } \{\langle 12\rangle,\langle 13\rangle,\ldots,\langle 1n\rangle\} \mbox{ has } k-2 \mbox{ sign flips}\,, \nonumber \\
&\mbox{the sequence } \{[ 12],[ 13],\ldots,[ 1n]\} \mbox{ has } k \mbox{ sign flips}\} \,,
\end{align}
where $s_{i,i+1,\ldots,i+j}$ are planar multiparticle Mandelstam variables: $s_{i,i+1,\ldots,i+j}=(p_i+p_{i+1}+\ldots+p_{i+j})^2$.
 Then the momentum amplituhedron $\mathcal{M}_{n,k}^{(\lambda,\widetilde\lambda)}$ directly in terms of kinematic data in the spinor helicity space is the intersection
$$
\mathcal{M}_{n,k}^{(\lambda,\widetilde\lambda)}:=\mathcal{V}_{n,k}\cap \mathcal{W}_{n,k}\,.
$$
There are various ways to compute the canonical form $\Omega_{n,k}$ on $\mathcal{M}_{n,k}^{(\lambda,\widetilde\lambda)}$, the most common including the introduction of auxiliary spaces \cite{Damgaard:2019ztj}, the Grassmannian integrals \cite{He:2018okq}, or the inverse-soft construction \cite{He:2018okq} (see also section \ref{sec:ISF}).
After finding the canonical form,
the amplitude can be extracted from it via the replacement 
\begin{equation}
A_{n,k}(\lambda, \tilde\lambda,\eta,\tilde\eta) = \delta^4(\lambda \tilde\eta+\tilde\lambda \eta)\,\Omega_{n,k}|_{d\lambda\rightarrow\eta,d\tilde\lambda\rightarrow\tilde\eta} \,.
\end{equation}
Finally, we notice that for the momentum amplituhedron $\mathcal{M}_{n,k}^{(\lambda,\tilde\lambda)}$, the degree of the canonical differential forms is independent of $k$ and equals $2n-4$. This allows us to write the superamplitude $\mathbb{A}_n$ as a single linear combination of forms for different $k$
\begin{equation}
\Omega_n = \sum_{k=2}^{n-2} \Omega_{n,k} \,,
\end{equation}
and therefore
\begin{equation}
\mathbb{A}_{n}(\lambda, \tilde\lambda,\eta,\tilde\eta) = \delta^4(\lambda \tilde\eta+\tilde\lambda \eta)\,\Omega_{n}|_{d\lambda\rightarrow\eta,d\tilde\lambda\rightarrow\tilde\eta} \,.
\end{equation}


\subsection{Kinematic Associahedron}
The kinematic associahedron $\mathcal{A}_n$ is the positive geometry whose canonical differential form gives tree-level amplitudes in massless bi-adjoint $\phi^3$ theory \cite{Arkani-Hamed:2017mur}. This is the theory of scalars carrying the adjoint representation of the product of two different color groups.
An $n$-point amplitude in this theory can be decomposed into double-partial amplitudes $m_n(\alpha | \beta)$, where  $\alpha$ and $\beta$  label  two color orderings, i.e.~a permutation of $n$ elements. In the following we will focus on the tree-level double-partial amplitudes with the same standard
ordering, $m^{(0)}_n(1,2,\ldots,n | 1,2,\ldots,n)=:m^{(0)}_n$. 
The amplitudes $m^{(0)}_n$ can be found by summing over all color-ordered trivalent planar graphs, each contributing the product of its propagators, and therefore  they are rational functions of Mandelstam variables.

The kinematic associahedron lives in the kinematic space $\mathcal{K}_n$ for $n$ massless particles. This space is linearly spanned by
 the Mandelstam variables $s_{i,j}$, which satisfy $n$ conditions of the form $\sum_{i\neq j} s_{i,j} = 0$. Therefore, its dimension is $\mathrm{dim} \,\mathcal{K}_n = \frac{n (n-3)}{2}$.
A natural choice for a basis of this space is given by the so-called planar variables:  given the standard ordering $(12\ldots n)$, one can define $\frac{n (n-3)}{2}$ variables 
\begin{equation}
\label{planarX}
{\bf X} = \{X_{i,j}\} := \{s_{i,i+1,\ldots,j-1}\}\,,
\end{equation}
which are Mandelstam variables formed of momenta of consecutive particles, and which can be visualised as the diagonals between vertices $i$ and $j$ of a convex $n$-gon. 

Similar to the momentum amplituhedron, the kinematic associahedron can be defined as the intersection of a positive region and an affine space.
The positive region  $\Delta_n$ is defined by the requirement that all planar variables $X_{i,j}$ are positive 
\begin{equation}
X_{i,j}\geq 0\,,  \quad  \, 1\leq i< j\leq n \,.
\end{equation}
This determines a top-dimensional cone inside $\mathcal{K}_n$. The affine subspace is the $(n-3)$-dimensional subspace $H_n \subset \mathcal{K}_n$ defined by requiring that 
\begin{equation}
c_{i,j} :=-s_{i,j} =X_{i,j} + X_{i+1,j+1} - X_{i,j+1} - X_{i+1,j} \,,
\end{equation}
are positive constants for all non-adjacent $1 \leq i<j < n$. 
Then the kinematic associahedron $\mathcal{A}_n$ is defined as: 
\begin{equation}
\mathcal{A}_n := \Delta_n \cap H_n \,.
\end{equation}
This is an $(n-3)$-dimensional subset of $\mathcal{K}_n$ which can be naturally parametrised by e.g. $X_{i,n}$ with $i=2,\ldots,n-2$. One can easily show that its boundary structure is identical to the $(n-3)$-dimensional associahedron \cite{Arkani-Hamed:2017mur}. 

The canonical form  of $\mathcal{A}_n$, which we call $\tilde\omega_n$,  can be found by using the fact that the associahedron is a simple polytope, i.e.~a $d$-dimensional polytope each of whose vertices are adjacent to exactly $d$ facets, see \cite{Arkani-Hamed:2017mur}. All facets are characterised by the vanishing of one of the planar variables and therefore we can write
\begin{equation}\label{eq:associahedronform}
\tilde\omega_n= \sum_{p=1}^{C_{n-2}} \mathrm{sign}(v_p) \bigwedge_{a=1}^{n-3}d\mathrm{log} X_{i_a,j_a} \,,
\end{equation}
where  $C_{n-2}$ is the number of vertices of the associahedron, i.e.~the Catalan number, and $\mathrm{sign}(v_p)$ are signs which can be fixed by requiring $\tilde\omega_n$ to be projective on $\mathcal{K}_n$.
The tree-level scattering amplitude $m_n^{(0)}$ is related to the differential form in the following way:
\begin{equation}
\label{eq:amplAssoc}
\tilde\omega_n=  m_n^{(0)}  d^{n-3}X\,.
\end{equation}


\section{Maps between Kinematic Spaces and Differential Forms}
\label{sec:maps}

In this section we present an extended discussion on various kinematic spaces which can be used to write the canonical differential forms of the momentum amplituhedron and associahedron, and on the maps which link them. We will then move to study how these maps induce an action on the differential forms, after which we will present our main result: the canonical differential forms of the momentum amplituhedron and the associahedron are related in a very specific (and surprising) way.

\subsection{Kinematic Spaces}
We start by presenting the various kinematic spaces in which we can write our canonical forms.

\paragraph{On-shell space $\mathcal{O}_{n}$.} This is the space where the momentum amplituhedron is defined: the bosonic part of the on-shell superspace which is parametrized by two $2\times n$ matrices, $(\lambda,\tilde\lambda)$, modulo momentum conservation which provides four relations between their entries.  Moreover, the canonical form of the momentum amplituhedron is written in terms of minors of these matrices, which reflects the $SL(2)\times SL(2)$ symmetry of the amplitude. Each $SL(2)$ symmetry reduces the dimension of the space by 3, and therefore the on-shell space $\mathcal{O}_n$ has dimension $\dim \mathcal{O}_n=4n-10$.

\paragraph{Little group invariant space $\mathcal{L}_n$.} 
The $\mathcal{N}=4$ sYM amplitude in non-chiral superspace is invariant under the little group scaling transformations
\begin{equation}\label{remove.scaling}
\lambda_i\to t_i\lambda_i\,,\qquad \qquad \tilde\lambda_i\to t_i^{-1}\tilde\lambda_i\,, \qquad \qquad\eta_i\to t_i\eta_i\,,\qquad \qquad \tilde\eta_i\to t_i^{-1}\tilde\eta_i\,.
\end{equation}
We can make this invariance manifest by parametrizing $(\lambda_i,\tilde\lambda_i)$ with a set of $n$ variables $t_i$ and calling the remaining variables collectively $\mathbf{a}$. There are exactly $3n-10$ independent $\mathbf{a}$ variables, and this comes from the fact that we start from a $(4n-10)$-dimensional on-shell space and remove $n$ variables $t_i$. 
In the following, we will use the same parametrization to rewrite the canonical form of the momentum amplituhedron.

 There are many ways in which we can parametrize the $(3n-10)$-dimensional space $\mathcal{L}_n$\footnote{We can think of the space $\mathcal{L}_n$ as the on-shell space $\mathcal{O}_{n}$ modulo the little group torus, $\mathcal{O}_{n}/T$, with $T=\mathbb{R}^n_+$, see also \cite{Arkani-Hamed:2019rds}.}. One important parametrization is obtained by first focusing on the $\lambda$ matrix and introducing variables $t_i$ as in \eqref{remove.scaling}, then the remaining $\lambda$-space has dimension $2n-3-n=n-3$. This is nothing else than the moduli space of a Riemann sphere with $n$ punctures and can be naturally parametrized using the Fock-Goncharov variables \cite{FGcoords}:
\begin{equation}\label{just.lambda}
\lambda =\left(\begin{matrix}
0&1&1&1&1&\ldots&1\\
-1&0&1&1+a_1&1+a_1+a_1a_2&\ldots&1+a_1+\ldots+a_1 a_2\ldots a_{n-3}
\end{matrix}\right) \,.
\end{equation} 
The $\tilde\lambda$ matrix can be parametrized by demanding that it is perpendicular to $\lambda$ and fixing the $SL(2)$ invariance. We will call this type of parametrization the {\it extended Fock-Goncharov parametrization}.

 In the extended Fock-Goncharov parametrization, by choosing the $\lambda$ matrix as the starting point we have broken the parity symmetry between $\lambda$ and $\tilde\lambda$. Replacing the roles of $\lambda$ and $\tilde\lambda$ would give us another parametrization in which $\tilde \lambda$ is written in terms of $n-3$ Fock-Goncharov variables and $\lambda$ depends on the remaining ones. This points to a natural set of coordinates which highlights the parity symmetry of amplitudes. Let us define the following cross-ratios:
\begin{equation}\label{crossratios}
R_{ijkl} = \frac{\langle ij\rangle \langle kl\rangle}{\langle il\rangle \langle jk\rangle} \quad,\quad \bar{R}_{ijkl} = \frac{[ij][kl]}{[il][jk]} \,.
\end{equation}
Note that $a_i$ in \eqref{just.lambda} can be written in terms of these cross-ratios as $a_i=R_{1i+1 i+2 i+3}$. There are $n-3$ algebraically independent cross-ratios $R$ and $n-3$ algebraically independent cross-ratios $\bar R$. In order to get the proper number of parameters for the little group invariant space $\mathcal{L}_n$, i.e.~$(3n-10)$, we need to supplement them with exactly $n-4$ Mandelstam variables. 
In this way, all variables are manifestly little group scaling invariant and parity symmetry is also manifest\footnote{However, in the extended Fock-Goncharov parametrization, cyclic symmetry is not manifest.}. We will give explicit expressions for the Fock-Goncharov parametrization when we study examples in the following sections.  

\paragraph{Space of Mandelstam variables $\mathcal{K}_n$ and Gram determinant surface $\mathcal{G}_n$.} As already mentioned, the associahedron is naturally realized in the kinematic space $\mathcal{K}_n$ of the scalar theory, i.e.~the space parametrized by the set of $\frac{n(n-3)}{2}$ planar Mandelstam variables. However, when the spacetime dimension $d$  is smaller than the number of independent massless momenta $n-1$, $d < n-1$,  not all planar Mandelstam variables are independent. There are indeed further constraints, the so-called {\it Gram determinant conditions}. 
For a fixed dimension $d$, Gram matrices are square $(d+1)\times (d+1)$ matrices which depend on $d+1$ momenta $p_i$:
\begin{equation}
G(p_{i_1},\ldots,p_{i_{d+1}}) = (s_{i,j})_{i,j \in \{i_1,i_2,\ldots,i_{d+1}\}}\,,
\end{equation}
and consist of two-particle Mandelstam variables $s_{i,j} = 2p_i\cdot p_j$, $i,j \in \{i_1,i_2,\ldots,i_{d+1}\}$. For $d<n-1$, the determinant of each Gram matrix must vanish, which imposes constraints on the planar variables.
Therefore, for the four-dimensional scalar theory one finds that the Gram determinant constraints start to appear from $n\geq 6$. Not all Gram determinant conditions are independent, and one can find that solving all of them reduces the number of independent parameters to $3n-10$, in agreement with the dimension of the little group scaling invariant space $\mathcal{L}_n$. 
We can therefore define a $(3n-10)$-dimensional space  $\mathcal{G}_n$ inside the $\frac{n(n-3)}{2}$-dimensional space of planar Mandelstam variables by imposing the Gram determinant conditions. We denote the coordinates on $\mathcal{G}_n$ by a collective label $\mathbf{x}$.
This will allow us to perform the push-forward of the associahedron canonical form to the surface where all Gram determinants vanish.
\bigskip

\subsection{Maps between Kinematic Spaces}
\label{subsec:maps}
\begin{figure}[t]
\begin{center}
\begin{tikzcd}
	\begin{tikzpicture}
		\node [fill=colourO!15,draw=colourO,rounded corners=10pt] {\begin{minipage}{2.3cm}
				\begin{center}
					$\mathcal{O}_n$\\
					on-shell space\\
					$(\lambda,\tilde\lambda)$\\
					$4n-10$\\
					{\color{red!90!colourO}$\Omega_{n,k}$}
		\end{center}\end{minipage}};
	\end{tikzpicture}
&\begin{tikzpicture}
	\node [fill=colourL!15,draw=colourL,rounded corners=10pt] {\begin{minipage}{2.51cm}
			\begin{center}
				$\mathcal{L}_n $\\
				LGS invariants\\
				$(\mathbf{a})$ or $(R,\bar R,s)$\\
				$3n-10$\\
				$\omega_{n,k},\omega_{n}$
	\end{center}\end{minipage}};
\end{tikzpicture}\arrow[l,"\mathbf{f}_n" above]\arrow[rr,bend right,"\mathbf{p}_n" below ]\arrow[r,"\mathbf{g}_n"]&\begin{tikzpicture}
\node [fill=colourG!15,draw=colourG,rounded corners=10pt] {\begin{minipage}{3.3cm}
		\begin{center}
			$\mathcal{G}_n$\\
			Mandelstams/Gram\\
			$(\mathbf{x})$\\
			$3n-10$\\
			$\nu_{n,k},\nu_{n}$
		\end{center}
\end{minipage}};
\end{tikzpicture}&\begin{tikzpicture}
\node [fill=colourK!15,draw=colourK,rounded corners=10pt] {\begin{minipage}{2.2cm}
		\begin{center}
			$\mathcal{K}_n$\\
			Mandelstams\\
			$(\mathbf{X})$ or $(\mathbf{s})$\\
			$\frac{n(n-3)}{2}$\\
			{\color{red!90!colourK}$\tilde\omega_{n}$}
\end{center}\end{minipage}};
\end{tikzpicture}\arrow[l,"\mathbf{h}_n" above]
\end{tikzcd}
\end{center}
\caption{Summary of the kinematic spaces and relations between them, together with the differential forms defined on these spaces. We highlight the canonical forms of the momentum amplituhedron $\Omega_{n,k}$ and of the associahedron $\tilde\omega_n$.}
\label{fig:maps}
\end{figure}
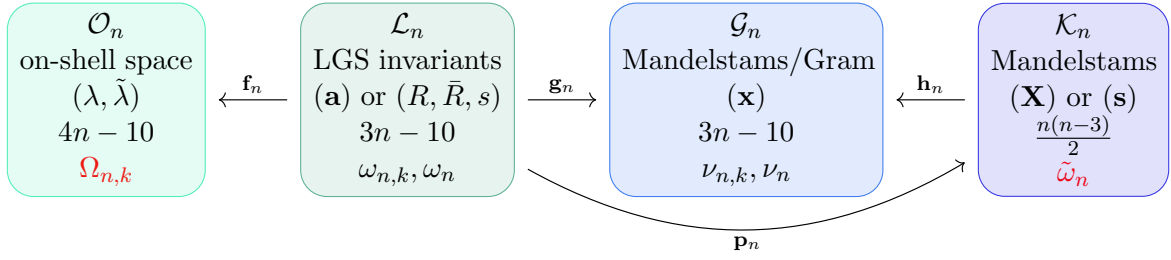

The kinematic spaces defined in the previous section are related among each other via maps, which we define in the following. We collect all these maps in Fig.~\ref{fig:maps}. In section \ref{sec:examples} we will give explicit expressions for these maps for the first few values of $n$. 

Removing the little group scaling dependence according to \eqref{remove.scaling} and parametrizing the on-shell space $\mathcal{O}_n$ by the little group scaling variables $t_i$  and the extended Fock-Goncharov variables $\mathbf{a}$ defines a function 
\begin{eqnarray}
\label{scaleAndFG}
{\mathbf{ f}}_n: \mathcal{L}_n \to \mathcal{O}_n\,,\qquad \underset{n+(3n-10)}{(t_i, {\bf a})} \mapsto  \underset{4n-10}{(\lambda,\tilde\lambda)}\,.
\end{eqnarray}
If we restrict to $t_i > 0$ then $\mathbf{f}_n$ is an invertible map. 

There is a natural map from the little group invariant space $\mathcal{L}_n$ to the Mandelstam space or its subset satisfying Gram determinant conditions,  $\mathcal{K}_n$ and $\mathcal{G}_n$, respectively, given by $s_{i,j}=\langle ij\rangle[ij]$. We denote the map to $\mathcal{G}_n$ by $\mathbf{g}_n$:
\begin{eqnarray}
\label{mapG}
{\bf g}_n  :\mathcal{L}_n\to \mathcal{G}_n\,,\qquad \underset{3n-10}{{\bf a}}\mapsto \underset{3n-10}{{\bf x}} \,,
\end{eqnarray}
and the map to $\mathcal{K}_n$ by $\mathbf{p}_n$:
\begin{eqnarray}
\label{mapP}
{\bf p}_n  :\mathcal{L}_n\to \mathcal{K}_n\,,\qquad \underset{3n-10}{{\bf a}}\mapsto \underset{\frac{n(n-3)}{2}}{{\bf X}} \,.
\end{eqnarray}
Importantly, these maps are rational maps and for $n\geq 5$ they are not invertible. Instead, one can find that the number of local inverses increases when one increases $n$. 

Finally, the Gram determinant conditions define a map from the space of all planar Mandelstam variables to its $(3n-10)$-dimensional subset:
\begin{equation}
\mathbf{h}_n:\mathcal{K}_n\to \mathcal{G}_n\,,\qquad \underset{\frac{n(n-3)}{2}}{{\bf X}}\mapsto \underset{3n-10}{{\bf x}} \,.
\end{equation}
Since this map is defined by imposing some number of Gram determinant constraints on $\mathbf{X}$, we are not able to write $\mathbf{h}_n$ explicitly. However, the solutions to these constraints define for us all possible inverse functions, which we can use to perform push-forwards of differential forms from $\mathcal{K}_n$ to $\mathcal{G}_n$, as explained in the following section.

\subsection{Differential Forms}

\paragraph{Definitions.}
The maps defined in section \ref{subsec:maps} can be used to relate the canonical forms of different positive geometries. We consider here two operations on differential forms: {\it pull-back} and {\it push-forward}. The pull-back of a differential form is a standard notion in differential geometry, while the push-forward of a differential form was discussed in \cite{Arkani-Hamed:2017tmz} for top forms. Below, we will extend their definition to differential forms which are not top-dimensional. 

If we consider a map  $\phi :A \to B$ and a form $\beta$ on  $B$, we can pull it back to a form $\alpha$ on $A$ using the map $\phi$. Let us assume that $A$ is an $n$-dimensional space with coordinates $(x_1,\ldots,x_n)$, $B$ is an $m$-dimensional space with coordinates $(y_1,\ldots,y_m)$, and let us write $\phi=(\phi_1,\ldots, \phi_m)$.
Then for any $k$-form $\beta$ written in the coordinate basis for $B$
\begin{equation}
\beta=\sum_{1\leq i_1\leq\ldots\leq i_k\leq m}\,\beta_{i_1\ldots i_k}(y_1,\ldots,y_m)dy_{i_1}\wedge \ldots \wedge dy_{i_k}\,,
\end{equation} 
the pull-back of $\beta$ is 
\begin{equation}
\alpha=\phi^*(\beta):= \sum_{1\leq i_1\leq\ldots\leq i_k\leq m}\beta_{i_1\ldots i_k}(\phi(x_1,\ldots,x_n))\, d\phi_{i_1}\wedge \ldots \wedge d\phi_{i_k}\,.\end{equation}
In practice, we simply substitute the explicit expressions for $y_i=\phi_i(x)$ into the form $\beta$.

Instead, if we want to start with a differential form on $A$ and use the map $\phi$ to find a corresponding form on $B$, then we can use the so-called push-forward defined in the following way. 
For a given point $b\in B$ we can find its pre-image, namely the collection of points $a_i$ in $A$ satisfying $\phi(a_i)=b$. Then there exists a neighbourhood $U_i$ of each point $a_i$ and a neighbourhood $V$ of $b$, such that we can define the inverse maps: $\psi_i=\phi|_{U_i}^{-1}:V\to U_i$. Then the push-forward of a form $\alpha$ on $A$ through $\phi$ is a differential form $\beta$ on $B$ given by the sum over all solutions of the pull-backs through the inverse maps $\psi_i$:
\begin{equation}
\beta=\phi_*\alpha=\sum_i\psi_i^*\alpha \,.
\end{equation}
In practice, we solve the equation $y=\phi(x)$ and for each solution $x=\psi_i(y)$ we perform the pull-back of $\alpha$ and then sum the resulting differential forms.

\paragraph{Comparing forms on $\mathcal{L}_n$.}
We  now   describe the action of the maps introduced above on the canonical forms of the momentum amplituhedron and the associahedron. Here we discuss their relation when we pull them back to the little group invariant space $\mathcal{L}_n$. Later on,  we will use push-forwards to compare these forms on the Gram determinant surface $\mathcal{G}_n$ inside the kinematic space $\mathcal{K}_n$.

Our first observation is that the pull-back of the momentum amplituhedron canonical form $\Omega_{n,k}$ through the map $\mathbf{f}_n$ defines a differential form on the little group invariant space $\mathcal{L}_n$. In particular, based on the examples we studied, we notice that the term with the highest degree in $dt_i$ is independent of $k$ as well as our choice of parametrization for $\mathcal{L}_n$. Explicitly, we find that
\begin{equation}\label{pullback.mom}
\mathbf{f}_n^{\ast}\,\Omega_{n,k}=\mu_n \wedge {\omega}_{n,k}+\mathcal{O}(d^{n-2}t)\,,
\end{equation}
where
\begin{equation}
\label{projmu}
\mu_n=\mu\left(\mathbb{P}^{n-1}\right) = \sum_{i=1}^{n} (-1)^{n-i} d\mathrm{log}t_1 \wedge \ldots\wedge \overline{d\mathrm{log}t_i}  \wedge \ldots\wedge d\mathrm{log}t_n \,,
\end{equation}
is the canonical form on the projective space $\mathbb{P}^{n-1}$, the overline indicates that that term is missing,  and $\mathcal{O}(d^{n-2}t)$ denotes terms which are of lower degree in $dt_i$. The explicit form of these lower-order terms depends on how we parametrize the space $\mathcal{L}_n$ and therefore they are not well-defined on $\mathcal{L}_n$. For this reason we will focus only on the top component $\mu_n\wedge \omega_{n,k}$ in \eqref{projmu}, and thus lose some information about the momentum amplituhedron form $\Omega_{n,k}$. Notice that since $\text{deg}\,\Omega_{n,k}=2n-4$ and $\text{deg}\,\mu_n=n-1$ then $\text{deg}\,\omega_{n,k}=n-3$. We call $\omega_{n,k}$ the {\it reduced momentum amplituhedron form}. Importantly, since this form is defined on $\mathcal{L}_n$, it is little group scaling invariant. In section \ref{sec:ISF} we will show how to construct it recursively using the inverse-soft construction.

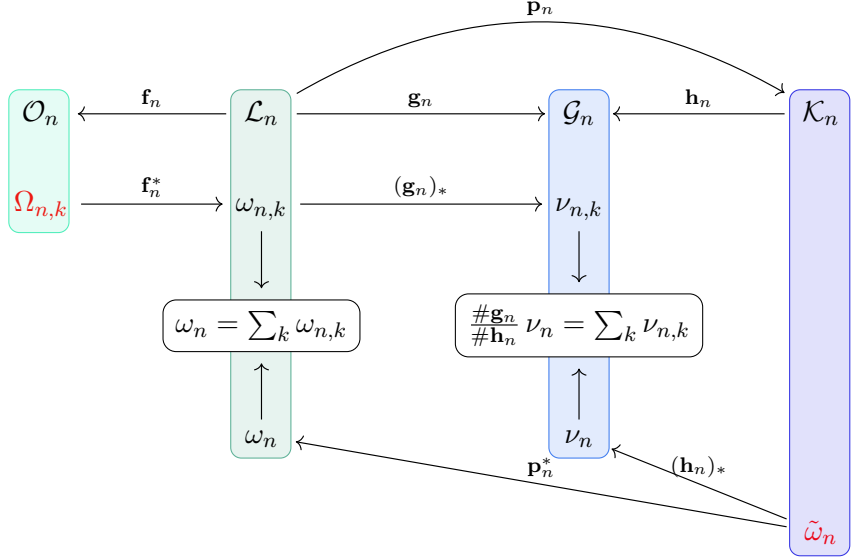
\begin{figure}[ht]
\begin{center}
\[
\tikz[ 
overlay]{
	\filldraw[fill=colourO!15,draw=colourO,rounded corners] (0.1,1.2) rectangle (0.9,3.1);
	\filldraw[fill=colourL!15,draw=colourL,rounded corners] (3.05,-1.8) rectangle (3.85,3.1);
	\filldraw[fill=colourG!15,draw=colourG,rounded corners] (7.28,-1.8) rectangle (8.08,3.1);
	\filldraw[fill=colourK!15,draw=colourK,rounded corners] (10.48,-3.1) rectangle (11.28,3.1);

}
\begin{tikzcd}
\,\mathcal{O}_n\,&\,\mathcal{L}_n\,\arrow[l,"\mathbf{f}_n" above]\arrow[rr,bend left,"\mathbf{p}_n" above ]\arrow[r,"\mathbf{g}_n"]&\,\mathcal{G}_n\,&\,\mathcal{K}_n\,\arrow[l,"\mathbf{h}_n" above]  \\  [-8mm]
 \phantom{\vdots} &  &  &  \\   [-8mm]
{\color{red!90!colourO}\Omega_{n,k}} \arrow[r, "\mathbf{f}_n^\ast"] 
& \omega_{n,k} \arrow[r, "(\mathbf{g}_n)_\ast"]\arrow{d}&\nu_{n,k}\arrow[d]  & \\
&\begin{tikzpicture}
	\node [fill=white,draw=black,rounded corners=5pt] {$\omega_n=\sum_{k}\omega_{n,k}$};
\end{tikzpicture}&\begin{tikzpicture}
\node [fill=white,draw=black,rounded corners=5pt] {$\frac{\# \mathbf{g}_n}{\#\mathbf{h}_n}\,\nu_n=\sum_{k}\nu_{n,k}$};
\end{tikzpicture} & \\
&\,\omega_n\,\arrow[u]
&\,\nu_n\,\arrow[u]&  \\
&&&{\color{red!90!colourK} \,\tilde\omega_n\,} \arrow[ul,"(\mathbf{h}_n)_\ast" above]\arrow[ull,"\mathbf{p}_n^\ast" above ]
\end{tikzcd}
\]
\end{center}
\caption{Relations between the kinematic spaces and the various differential forms on them defined in this section. We highlight the canonical forms of the momentum amplituhedron $\Omega_{n,k}$ and of the associahedron $\tilde\omega_n$.}
\label{diff.forms.diag}
\end{figure}

Starting from the opposite end of our story, we can pull the associahedron form $\tilde\omega_n$  back to the little group invariant space $\mathcal{L}_n$ using the map $\mathbf{p}_n$ to define
\begin{equation}
\omega_{n}=\mathbf{p}_n^\ast \,\tilde\omega_n \,.
\end{equation}
We find that this pull-back of the associahedron form to $\mathcal{L}_n$ is equal to the sum over all $k$-sectors of the reduced momentum amplituhedron form: 
\begin{align}
\label{sum.omega}
\omega_{n}=\sum_{k=2}^{n-2}\omega_{n,k}\,.
\end{align}
Therefore, we can write that on $\mathcal{L}_n$ the (pull-back of the) differential form for the full $\mathcal{N}=4$ sYM amplitude $\Omega_{n} $ and the (pull-back of the) differential form for the bi-adjoint $\phi^3$ amplitude $\tilde\omega_n$ are related in the following way:
\begin{equation}
\Omega_{n} = \sum_{k=2}^{n-2} \Omega_{n,k} \overset{\mathbf{f}^*_n}{\longrightarrow} \mu_n \wedge \sum_{k=2}^{n-2} {\omega}_{n,k} =  \mu_n \wedge {\omega}_{n}  
\overset{\mathbf{p}^*_n}{\longleftarrow} \mu_n \wedge \tilde\omega_n
\,.
\end{equation}
This relation is illustrated in diagrammatic form in Fig.~\ref{diff.forms.diag} and  is the main result of this paper.
As it will be more evident in the explicit examples in section \ref{sec:examples}, \eqref{sum.omega} relates the singularity structure of the momentum amplituhedron  to that of the kinematic associahedron. In particular, the factorization channels given by the vanishing of planar Mandelstam variables are the same.

\paragraph{Comparing forms on $\mathcal{G}_n$.}
The form $\omega_{n,k}$ in \eqref{pullback.mom} can be further pushed forward using the map $\mathbf{g}_n$ to define
\begin{equation}
\nu_{n,k}=(\mathbf{g}_n)_\ast\, \omega_{n,k} \,.
\end{equation}
On the other hand, the Gram determinant map $\mathbf{h}_n$ allows us to also push the associahedron canonical form $\tilde\omega_n$ forward onto the Gram determinant surface to define
\begin{equation}
\nu_{n}=(\mathbf{h}_n)_\ast \,\tilde\omega_n \,.
\end{equation}
This relation is again illustrated in Fig.~\ref{diff.forms.diag}. As we will present more explicitly later, in the various examples we checked, the following statement holds true \begin{equation}\label{sum.nu}
\sum_{k=2}^{n-2}\nu_{n,k}=\begin{cases}\nu_n,&\text{for }n=4,\\2\nu_n,&\text{for }n>4.\end{cases}
\end{equation}
We found that the intriguing factor of 2 on the right hand side of \eqref{sum.nu} can be explained in the following way. Let $\#\mathbf{g}_n$ (resp.\ $\#\mathbf{h}_n$) be the degree of the map $\mathbf{g}_n$ (resp.\ $\mathbf{h}_n$), i.e.~the number of solutions to the equation $y=\mathbf{g}_n(x)$ (resp.\ $y=\mathbf{h}_n(x)$).  Then the formula \eqref{sum.nu} can be rewritten as
\begin{align}\label{sum.nu.ratio}
	\sum_{k=2}^{n-2}\nu_{n,k}=\frac{\#\mathbf{g}_n}{\#\mathbf{h}_n}\nu_n.
\end{align}
This can be explicitly checked for $n=4,5,6,7$, for which we get\footnote{Depending on the choice of basis for $\mathcal{G}_7$ we also find $(\#\mathbf{g}_7,\#\mathbf{h}_7)=(16,8)$, but their ratio is still 2.}: $(\#\mathbf{g}_4,\#\mathbf{h}_4)=(1,1)$, $(\#\mathbf{g}_5,\#\mathbf{h}_5)=(2,1)$, $(\#\mathbf{g}_6,\#\mathbf{h}_6)=(4,2)$ and $(\#\mathbf{g}_7,\#\mathbf{h}_7)=(8,4)$. We believe that this pattern extends beyond $n=7$.
We postpone a more detailed  discussion on \eqref{sum.nu.ratio} until the next section when we consider explicit examples.


\section{Examples}
\label{sec:examples}

To illustrate the relations we provided in section \ref{sec:maps}, we now present a few examples for small values of $n$. In particular, we give some explicit expressions for the maps and differential forms introduced there.  

\subsection{Four-point Amplitudes}
The simplest case in which the momentum amplituhedron and kinematic associahedron are non-trivial is for four-particle scattering. For the kinematic associahedron $\mathcal{A}_4$ defined on $\mathcal{K}_4$, the canonical form is \cite{Arkani-Hamed:2017mur}
\begin{equation}
 \label{eq.associaForm4}
 \tilde{\omega}_4 = d \log \frac{X_{1,3}}{X_{2,4}} = d\log  \frac{s_{1,2}}{s_{2,3}} \,.
\end{equation}
On the other hand, the $\mathcal{N}=4$ sYM amplitude consists of only one $k$-sector, namely $k=2$, and we need to consider just one momentum amplituhedron geometry $\mathcal{M}_{4,2}^{(\lambda,\tilde\lambda)}$, for which the canonical form written in on-shell space $\mathcal{O}_4$ is \cite{He:2018okq}
\begin{equation}
\label{eq.momamp.4}
\Omega_{4,2} = d\mathrm{log}\frac{\langle 12 \rangle}{\langle 13\rangle} \wedge d\mathrm{log}\frac{\langle 23\rangle}{\langle 13\rangle} \wedge d\mathrm{log}\frac{\langle 34\rangle}{\langle 13\rangle} \wedge d\mathrm{log}\frac{\langle 14 \rangle}{\langle 13 \rangle}  \,.
\end{equation}

\paragraph{Maps.}
In order to define the map $\mathbf{f}_4: \mathcal{L}_4\to \mathcal{O}_4$ we remove the little group scaling and parametrize $\lambda$ and $\tilde{\lambda}$ using the extended Fock-Goncharov variables which read
\begin{equation}\label{lambda4}
\lambda = 
\begin{pmatrix}
0 & t_2 & t_3 & t_4\\
-t_1 & 0 & t_3 &t_4( 1+a_1)
\end{pmatrix} \,, \quad \tilde\lambda = \begin{pmatrix}
t_1^{-1}a_2 & -t_2^{-1}a_2 &t_3^{-1} a_2 & 0\\
t_1^{-1}(1+a_1) & -t_2^{-1} & 0 & t_4^{-1}
\end{pmatrix} \,,
\end{equation}
where we have made a particular choice of $SL(2)$ when parametrizing $\tilde{\lambda}$. On the other hand we can define a map $\mathbf{p}_4:\mathcal{L}_4\to \mathcal{K}_4$ which takes the form:
\begin{equation}
 \mathbf{p}_4: \, s_{1,2} =\langle 12 \rangle [12] = a_1 a_2, \quad s_{2,3} = \langle 23 \rangle [23] = a_2.
\end{equation}

Finally, since there are no Gram determinant constraints for $n=4$, one finds that $\mathcal{G}_4=\mathcal{K}_4$ and $\mathbf{h}_4=\mathbf{I}_4$ is the identity map, which implies that $\mathbf{g}_4 = \mathbf{p}_4$ and $\nu_4=\tilde\omega_4$.

\paragraph{Comparing forms on $\mathcal{L}_4$.} Using the parametrization \eqref{lambda4}, we can now pull $\Omega_{4,2}$ back to the $\mathcal{L}_4$ space yielding
\begin{equation}
	\mathbf{f}_4^\ast\,\Omega_{4,2} = 
	\mu_4 \wedge d\mathrm{log}a_1\,\,\longrightarrow\,\, \omega_{4,2}=d\mathrm{log}a_1=d\mathrm{log}R_{1234}\,,
\end{equation}
with $\mu_4$ defined in \eqref{projmu} and the cross-ratio $R_{1234}$ defined in \eqref{crossratios}. We compare it to the pull-back of $\tilde\omega_4$ using the map $\mathbf{p}_4$ to get again
\begin{equation}
\omega_4=\mathbf{p}_4^\ast\,\tilde\omega_4=d\mathrm{log}R_{1234} \,.
\end{equation}
Therefore on $\mathcal{L}_4$ we have
\begin{equation}
\omega_4 = \omega_{4,2} \,.
\end{equation}

\paragraph{Comparing forms on $\mathcal{G}_4=\mathcal{K}_4$.} Since the map $\mathbf{p}_4$ is invertible, pushing the form $\omega_{4,2}$ forward via $\mathbf{p}_4$ returns the associahedron form
\begin{equation}\label{momtoass4}
\nu_{4,2}=(\mathbf{p}_4)_\ast\, d\mathrm{log}R_{1234}=  d\mathrm{log}\frac{s_{1,2}}{s_{2,3}}\, =  \nu_4 \,.
\end{equation}

\subsection{Five-point Amplitudes}
The canonical form for the associahedron in kinematic space $\mathcal{K}_5$ is \cite{Arkani-Hamed:2017mur}
\begin{align}\label{tildeomega5}
\tilde\omega_5&=d\mathrm{log}\frac{X_{1,3}}{X_{2,4}}\wedge d\mathrm{log}\frac{X_{1,3}}{X_{1,4}}+d\mathrm{log}\frac{X_{1,3}}{X_{2,5}}\wedge d\mathrm{log}\frac{X_{3,5}}{X_{2,4}}.
\end{align}
From the momentum amplituhedron side, there are two geometries contributing to the superamplitude $\mathbb{A}_5$ coming from the $ k=2$ and $k=3$ sectors. Their canonical forms are given by
\begin{align}
	\label{eq.momamp.5}
	\Omega_{5,2} &= -d\log\frac{\langle13\rangle}{\langle14\rangle}\wedge d\log\frac{\langle34\rangle}{\langle14\rangle}\wedge d\log\frac{\langle45\rangle}{\langle14\rangle}\wedge d\log\frac{\langle51\rangle}{\langle14\rangle}\wedge d\log\frac{\langle12\rangle}{\langle13\rangle}\wedge d\log\frac{\langle23\rangle}{\langle13\rangle},
\end{align}
and $\Omega_{5,3}$ can be found from $\Omega_{5,2}$ by replacing $ \langle \rangle \rightarrow [] $. 

\paragraph{Maps.}
We define the map $\mathbf{f}_5:\mathcal{L}_5\to\mathcal{O}_5$ by removing the little group scaling and choosing an extended Fock-Goncharov parametrization on $\mathcal{L}_5$ which depends on $3 \times 5 -10 = 5 $ variables. We choose the parameters $\{ R_{1234}, R_{1345}, \bar{R}_{1234}, \bar{R}_{1345}, s_{1,2}\} $, which allow us to write $\lambda$ as
 \begin{align}\label{lambda5}
 \lambda&=\left(\begin{matrix}
 0&t_2&t_3&t_4&t_5\\
 -t_1&0&t_3&t_4(1+R_{1234})&t_5(1+R_{1234}+R_{1234}R_{1345})
 \end{matrix}\right) \,,
 \end{align}
 and the parametrization of $\tilde\lambda$ can be found in Appendix \ref{app:tildelambda}. 
 On the other hand, the map $\mathbf{p}_5:\mathcal{L}_5\to\mathcal{K}_5$ can be easily found by calculating the minors of the matrices \eqref{lambda5} and \eqref{tildelambda5}. The map $\mathbf{p}_5$ is a rational map and it is not invertible. Instead, we can find two local inverses which take the form
  \begin{align}\label{p5inverse}
  \mathbf{p}_{5,\pm}^{-1}:\qquad \begin{matrix}	R_{1234}&= \frac{s_{1,2} \,s_{2,3}+s_{3,4} \, s_{2,3}-s_{3,4}\, s_{4,5}-s_{1,2} \, s_{5,1}+s_{4,5}\, s_{5,1}\pm\sqrt{\Delta}}{2 s_{2,3} (s_{2,3}-s_{4,5}-s_{5,1})}\\
 	R_{1345}&= \frac{s_{1,2}\, s_{2,3}-s_{3,4}\, s_{2,3}+s_{3,4}\, s_{4,5}-s_{1,2}\, s_{5,1}+s_{4,5}\, s_{5,1}\mp\sqrt{\Delta}}{2 s_{3,4}\, s_{5,1}}\\
 	\bar{R}_{1234}&= \frac{s_{1,2}\, s_{2,3}+s_{3,4} \, s_{2,3}-s_{3,4}\, s_{4,5}-s_{1,2}\, s_{5,1}+s_{4,5}\, s_{5,1}\mp\sqrt{\Delta}}{2 s_{2,3} (s_{2,3}-s_{4,5}-s_{5,1})}\\
 	\bar{R}_{1345}&= \frac{s_{1,2} \,s_{2,3}-s_{3,4}\, s_{2,3}+s_{3,4}\, s_{4,5}-s_{1,2}\, s_{5,1}+s_{4,5}\, s_{5,1}\pm \sqrt{\Delta}}{2 s_{3,4}\, s_{5,1}}\end{matrix}\quad,
 \end{align}
where the two solutions are distinguished by the sign in front of the square root of
\begin{equation}
\Delta=(s_{2,3}\, s_{3,4}+s_{1,2}
	(s_{2,3}-s_{5,1})+s_{4,5} (s_{5,1}-s_{3,4}))^2-4 s_{1,2}\, s_{2,3}\, s_{3,4} (s_{2,3}-s_{4,5}-s_{5,1}) \,.
\end{equation}
Interestingly, we note that the conjugation operation, interchanging $R$ and $\bar{R}$, exchanges the sign in front of the $\sqrt{\Delta}$:
\begin{equation}
	\bar{R}_{i,j,k,l} = R_{i,j,k,l}\vert_{\sqrt{\Delta} \rightarrow -\sqrt{\Delta}} \,.
\end{equation}

As for $n=4$, no Gram determinant conditions arise for $n=5$ and we have that $ \mathcal{G}_5 =\mathcal{K}_5$, $\mathbf{g}_5 = \mathbf{p}_5$, and $ \mathbf{h}_5 = \mathbf{I}_5$, the identity map. This implies that $\nu_5=\tilde\omega_5$.
 
 \paragraph{Comparing forms on $\mathcal{L}_5$.}
 Pulling back the momentum amplituhedron canonical forms to $\mathcal{L}_5$ we get 
 \begin{align}
 	&\mathbf{f}_5^\ast \Omega_{5,2} = \mu_5 \wedge R_{1234} \wedge R_{1345} &\longrightarrow \qquad\omega_{5,2} = d \log R_{1234} \wedge R_{1345} \,,\\
 	&\mathbf{f}_5^\ast \Omega_{5,3} = \mu_5 \wedge \bar{R}_{1234} \wedge \bar{R}_{1345}+\mathcal{O}(d^{3}t) &\longrightarrow \qquad \omega_{5,3} = d\log \bar{R}_{1234} \wedge \bar{R}_{1345}  \,.
 \end{align}
 We note that the pull-back of $\Omega_{5,3}$ contains several terms which are of lower degree in $dt_i$'s. Their explicit form is however not needed in our definition \eqref{pullback.mom} of the reduced forms $\omega_{5,k}$.
 
Starting from the kinematic associahedron $\mathcal{A}_5$, we can pull the associahedron form \eqref{tildeomega5} back to the $\mathcal{L}_5$ space, yielding
\begin{equation}
\omega_5=\mathbf{p}_5^\ast\, \tilde\omega_5=d \log R_{1234} \wedge R_{1345}+d\log \bar{R}_{1234} \wedge \bar{R}_{1345} =\omega_{5,2}+\omega_{5,3} \,.
\end{equation}
This verifies our first main statement \eqref{sum.omega}.
 
 \paragraph{Comparing forms on $\mathcal{G}_5=\mathcal{K}_5$.}
As there is no non-trivial Gram determinant condition, we can push $\omega_{5,k}$  directly to the native space of the associahedron $\mathcal{K}_5$ via the map $\mathbf{p}_5$. To do this, we need to pull $\omega_{5,k}$ back using the two functions \eqref{p5inverse} and then sum the resulting differential forms. Importantly, the square roots present in these inverse functions cancel out in the sum and we get
\begin{equation}
\nu_{5,2}=(\mathbf{p}_5)_\ast \,\omega_{5,2}=\tilde{\omega}_5\,,\qquad \nu_{5,3}=(\mathbf{p}_5)_\ast\, \omega_{5,3}=\tilde{\omega}_5\,.
\end{equation}
This verifies our second main statement \eqref{sum.nu}, since we have 
\begin{equation}
\nu_{5,2} + \nu_{5,3} =  2 \tilde\omega_5 = 2\nu_5\,.
\end{equation}

\subsection{Six-point Amplitudes}
For the $n=6$ case, the canonical form for the associahedron $\mathcal{A}_6$ on $\mathcal{K}_6$ is \cite{Arkani-Hamed:2017mur}
\begin{align}\label{tildeomega6}
\tilde{\omega}_6 = d \log \frac{X_{2,4}}{X_{1,3}} \wedge d \log \frac{X_{1,4}}{X_{4,6}} \wedge d\log \frac{X_{1,5}}{X_{4,6}} + d \log \frac{X_{2,6}}{X_{13}} \wedge d \log \frac{X_{3,6}}{X_{1,3}} \wedge d\log \frac{X_{4,6}}{X_{3,5}}\nonumber -\\
- d \log \frac{X_{2,6}}{X_{1,5}} \wedge d \log \frac{X_{2,5}}{X_{3,5}} \wedge d \log \frac{X_{2,4}}{X_{3,5}} + d \log \frac{X_{2,4}}{X_{1,3}} \wedge d \log \frac{X_{4,6}}{X_{3,5}}\wedge d\log \frac{X_{2,6}}{X_{1,5}} \,.
\end{align}
For the momentum amplituhedron we have three different sectors which contribute to the superamplitude $\mathbb{A}_6$, namely $k=2$, $k=3$, and $k=4$. The differential form for the momentum amplituhedron $\mathcal{M}_{6,2}^{(\lambda,\tilde\lambda)}$ on the $\mathcal{O}_6$ space can be written as
\begin{align}
	\nonumber \Omega_{6,2} &= -d\log\frac{\langle14\rangle}{\langle15\rangle}\wedge d\log\frac{\langle45\rangle}{\langle15\rangle}\wedge d\log\frac{\langle56\rangle}{\langle15\rangle}\wedge d\log\frac{\langle61\rangle}{\langle15\rangle}\wedge d\log\frac{\langle13\rangle}{\langle14\rangle}\wedge d\log\frac{\langle34\rangle}{\langle14\rangle}\\
	&\hspace{2.4cm}\wedge d\log\frac{\langle12\rangle}{\langle13\rangle}\wedge d\log\frac{\langle23\rangle}{\langle13\rangle} \,,
	\label{eq.momamp.6}
\end{align}
and again one can find the answer for the $k=4 $ sector through the conjugation operation, $\langle \rangle  \rightarrow []$. 
While the $\text{MHV}$ and $\overline{\text{MHV}}$  canonical forms are rather simple to write down, the $k=3$ differential form is more involved since it is written as a sum of three Britto-Cachazo-Feng-Witten (BCFW) terms \cite{Britto:2004ap,Britto:2005fq}. For their explicit expression see \cite{He:2018okq} or section \ref{sec:ISF} where we recall how to construct BCFW differential forms using the inverse-soft construction. Here we just recall that
\begin{align}
 \Omega_{6,3} &= \Omega_{6,3}^{\gamma_2}  + \Omega_{6,3}^{\gamma_4}  + \Omega_{6,3}^{\gamma_6} = \Omega_{6,3}^{\gamma_1}  + \Omega_{6,3}^{\gamma_3}  + \Omega_{6,3}^{\gamma_5}  \,, 
 \label{BCFW6}
\end{align}
where $\Omega_{6,3}^{\gamma_i} $ indicates the BCFW term with vanishing minor $\gamma_i :=(i, i+1,i+2)$.

\paragraph{Maps.}
In order to define the map $\mathbf{f}_6:\mathcal{L}_6\to \mathcal{O}_6$ we again use the extended Fock-Goncharov parametrization for $\lambda$ and $\tilde\lambda$, choosing as our $3\times 6-10=8$ parameters the cross-ratios $\{R_{1234},R_{1345},R_{1456}\}$, 
the cross-ratios $\{\bar{R}_{1234},\bar{R}_{1345},\bar{R}_{1456}\}$, and two Mandelstam variables, say $\{s_{1,2},s_{2,3}\}$. The explicit expressions for $\lambda$ can be read off from \eqref{just.lambda} but the ones for $\tilde\lambda$ become very large and we will not include them here.

Using the extended Fock-Goncharov parametrization one can easily construct the maps $\mathbf{g}_6:\mathcal{L}_6\to \mathcal{G}_6$ and $\mathbf{p}_6:\mathcal{L}_6\to \mathcal{K}_6$ using $s_{i,j}=\langle ij\rangle[ij]$ and substituting the explicit forms of matrices $\lambda$ and $\tilde\lambda$. Here, we need to decide which planar Mandelstam variables we use to parametrize the space $\mathcal{G}_6$ and our choice is: $(s_{1,2},s_{2,3},s_{3,4},s_{4,5},s_{5,6},s_{6,1},s_{1,2,3},s_{2,3,4})$. For the push-forward $(\mathbf{g}_6)_\ast$ we need to invert the map $\mathbf{g}_6$ which leads to four solutions, one of which is
\begin{align}
\label{p6inverse}
\mathbf{g}_{6,1}^{-1}:\qquad
\begin{tabular}{ll}$R^{(1)}_{1234}=$&$\frac{s_{1,2} s_{2,3}+s_{2,3} s_{3,4}-s_{2,3} s_{5,6}-s_{3,4} s_{1,2,3}-s_{1,2} s_{2,3,4}+s_{1,2,3} s_{2,3,4}-\sqrt{\Delta_1}}{2 s_{2,3}
   (s_{2,3}+s_{5,6}-s_{1,2,3}-s_{2,3,4})}$\\
$R^{(1)}_{1345}=$ & $  \frac{-s_{1,2} s_{2,3}+s_{2,3} s_{3,4}+s_{2,3} s_{5,6}-s_{3,4} s_{1,2,3}+s_{1,2} s_{2,3,4}-s_{1,2,3} s_{2,3,4}-\sqrt{\Delta_1}
  }{2 s_{3,4}
   (s_{5,6}+s_{6,1}-s_{2,3,4}) }\times$\\
   &$\times\frac{ s_{2,3} s_{5,6}+s_{4,5} s_{5,6}-s_{5,6} s_{6,1}+s_{6,1} s_{1,2,3}-s_{4,5} s_{2,3,4}-s_{1,2,3} s_{2,3,4}-\sqrt{\Delta_2}}{2(s_{2,3} s_{5,6}-s_{1,2,3} s_{2,3,4})}
  $\\
  $R^{(1)}_{1456}=$& $\frac{-s_{2,3} s_{5,6}+s_{4,5} s_{5,6}+s_{5,6} s_{6,1}-s_{6,1} s_{1,2,3}-s_{4,5} s_{2,3,4}+s_{1,2,3} s_{2,3,4}-\sqrt{\Delta_2}}{2 s_{4,5}
   s_{6,1}} $
   \end{tabular}
\end{align}
and 
\begin{equation}
\bar{R}^{(1)}_{ijkl}=R^{(1)}_{ijkl}|_{\sqrt{\Delta_1}\leftrightarrow -\sqrt{\Delta_1},\sqrt{\Delta_2}\leftrightarrow -\sqrt{\Delta_2}}\,,
\end{equation}
where the arguments of the square roots, $\Delta_1$ and $\Delta_2$, are written explicitly in App.~\ref{app:n6}.
The remaining three inverses of $\mathbf{g}_6$ can be obtained by exchanging the signs in front of the square roots
\begin{equation}
R^{(2)}_{ijkl}=R^{(1)}_{ijkl}|_{\sqrt{\Delta_1}\leftrightarrow -\sqrt{\Delta_1}}\,,\quad R^{(3)}_{ijkl}=R^{(1)}_{ijkl}|_{\sqrt{\Delta_2}\leftrightarrow -\sqrt{\Delta_2}}\,,\quad R^{(4)}_{ijkl}=R^{(1)}_{ijkl}|_{\sqrt{\Delta_1}\leftrightarrow -\sqrt{\Delta_1},\sqrt{\Delta_2}\leftrightarrow -\sqrt{\Delta_2}} \,.
\end{equation}

To define the map $\mathbf{h}_6$, we notice that the kinematic space $\mathcal{K}_6$ is nine-dimensional but the planar Mandelstam variables satisfy one Gram condition in four dimensions, reducing it to the eight-dimensional $\mathcal{G}_6$ space. When solving the Gram determinant condition one needs to decide with respect to which variables one wants to solve it: we decided to solve for $s_{3,4,5}$ and therefore parametrize the $\mathcal{G}_6$ as above.
Then $s_{3,4,5}$ can be found by solving the Gram determinant condition to find two solutions
	\begin{align}\label{Gram6}
	\mathbf{h}_{6,\pm}^{-1}:\qquad	s_{3,4,5}=\frac{\Gamma\pm\sqrt{\Delta_1\Delta_2}}{2s_{1,4}Q}\,,
	\end{align}
where $\Delta_1$ are $\Delta_2$ are the same as before, $s_{1,4}=s_{2,3} + s_{5,6} - s_{1,2,3} - s_{2,3,4}$ and $Q=s_{2,3} s_{5,6} - 
s_{1,2,3} s_{2,3,4}$, and the explicit form for $\Gamma$ can be found in App.~\ref{app:n6}.

\paragraph{Comparing forms on $\mathcal{L}_6$.} For the $\text{MHV}/\overline{\text{MHV}}$ sectors the pull-back of $\Omega_{6,2}$ and $\Omega_{6,4}$ on $\mathcal{L}_6$ takes a very simple form
\begin{align}
&\mathbf{f}^\ast_6 \Omega_{6,2} = \mu_6 \wedge \omega_{6,2}  &\longrightarrow && 
\omega_{6,2}=d\mathrm{log}R_{1234}\wedge d\mathrm{log}R_{1345}\wedge d\mathrm{log}R_{1456} \label{eq:w62}\,,\\
&\mathbf{f}^\ast_6 \Omega_{6,4} = \mu_6 \wedge \omega_{6,4}+ \mathcal{O}(d^4t) &\longrightarrow && \omega_{6,4}=d\mathrm{log}\bar{R}_{1234}\wedge d\mathrm{log}\bar{R}_{1345}\wedge d\mathrm{log}\bar{R}_{1456}\,.
\label{eq:w64}
\end{align}
For $k=3$, we can use the inverse-soft construction (see section \ref{sec:ISF}) to find the following compact expression for $\omega_{6,3}$ 
\begin{align}
	\nonumber &\mathbf{f}^\ast_6 \Omega_{6,3} = \mu_6 \wedge \omega_{6,3} +\mathcal{O}(d^4t)  &\longrightarrow&& \omega_{6,3}&=d\log R_{561\hat{2}}^{(234)}\wedge d\log R_{15\hat{2}\hat{4}}^{(234)}\wedge d\log R_{1\hat{2}\hat{3}\hat{4}}^{(234)}\\
	\nonumber &&&&&+d\log R_{123\hat{6}}^{(456)}\wedge d\log R_{31\hat{4}\hat{6}}^{(456)}\wedge d\log R_{1\hat{4}\hat{5}\hat{6}}^{(456)}\\
	&&&&&+d\log R_{345\hat{6}}^{(612)}\wedge d\log R_{54\hat{6}\hat{2}}^{(612)}\wedge d\log R_{5\hat{6}\hat{1}\hat{2}}^{(612)},\label{eq:w63}
\end{align}
where $R_{ijkl}^{\gamma}$ is the standard invariant cross-ratio built out of angle brackets, see \eqref{BCFW6} for an explanation on the label $\gamma$, and hatted particles $\hat{i}$ in $R^{\gamma}$ are defined as
$ \lambda^{\alpha}_{\hat{i}}=\sum_{j\in\gamma} \lambda^{\alpha}_j [ji]$. 
Finally, when we pull the associahedron form \eqref{tildeomega6} back to $\mathcal{L}_6$ using the map $\mathbf{p}_6$ we find that 
\begin{equation}
\omega_6=\mathbf{p}_6^\ast \, \tilde\omega_6=\omega_{6,2}+\omega_{6,3}+\omega_{6,4}\,.\label{eq:w6}
\end{equation}

\paragraph{Comparing forms on $\mathcal{G}_6$.}
This is the first time when the space $\mathcal{G}_n$ differs from $\mathcal{K}_n$ and we can perform a non-trivial push-forward of the associahedron form $\tilde\omega_n$ to $\mathcal{G}_n$. In order to do that we sum pull-backs of $\tilde\omega_6$ using the two solutions to the Gram determinant condition \eqref{Gram6} and we find
\begin{equation}
\nu_6= (\mathbf{h}_6)_\ast \tilde\omega_{6}\,.
\end{equation}
On the other hand we can use the four inverses of $\mathbf{g}_6$ in \eqref{p6inverse} to find 
\begin{equation}
	\nu_{6,k} = (\mathbf{g}_6)_\ast \, \omega_{6,k}\,\qquad k=2,3,4\,.\\
\end{equation}
One intriguing observation is that the differential forms $\nu_{6,k}$ behave non-canonically, with residues on zero-dimensional boundaries no longer restricted to $\pm 1$. A more detailed discussion on this surprising behaviour for the case $\nu_{6,2}$, as well as its explicit form, can be found in appendix \ref{app:nu62}. 

Finally, we have explicitly checked that
\begin{equation}\label{eq:sum.nu.6}
	 \nu_{6,2} + \nu_{6,3} + \nu_{6,4} = 2 \nu_6\,.
\end{equation}

The appearance of the factor of $2$ above is a property of the push-forwards, not of the specific differential forms involved. To see this, we start by rewriting \eqref{eq:sum.nu.6} as 
\begin{align}
	(\mathbf{g}_{6})_\ast\mathbf{p}_6^\ast\,\tilde{\omega}_6=2(\mathbf{h}_{6})_\ast\tilde{\omega}_6\,.
\end{align}
Recall that $\mathcal{G}_6$ is obtained from $\mathcal{K}_6$ by solving the Gram determinant condition with respect to $s_{3,4,5}$. If we take $\beta$ to be an arbitrary differential form on $\mathcal{K}_6$ which does not depend on $s_{3,4,5}$, then
\begin{align} \label{eq:beta-h}
(\mathbf{h}_{6})_\ast\,\beta=\#\mathbf{h}_{6}\,\beta,
\end{align}
and
\begin{align} \label{eq:beta-g}
	(\mathbf{g}_{6})_\ast\mathbf{p}_{6}^\ast\,\beta=\#\mathbf{g}_{6}\,\beta,
\end{align}
follow trivially, where $\#\mathbf{g}_{6}=4$ (resp.\ $\#\mathbf{h}_{6}=2$) counts the degree of $\mathbf{g}_{6}$ (resp.\ $\mathbf{h}_{6}$). Combining \eqref{eq:beta-h} and \eqref{eq:beta-g} we have
\begin{align}\label{eq:ratio.6}
	(\mathbf{g}_{6})_\ast\mathbf{p}_{6}^\ast\, \beta=\frac{\#\mathbf{g}_{6}}{\#\mathbf{h}_{6}} (\mathbf{h}_{6})_\ast\,\beta=2(\mathbf{h}_{6})_\ast\,\beta.
\end{align}
To see that \eqref{eq:ratio.6} holds for all differential forms on $\mathcal{K}_6$, including those which do depend on $s_{3,4,5}$, observe that the four solutions for $\mathbf{g}_{6,i}^{-1}$ are related by a $\mathbb{Z}_2\times\mathbb{Z}_2$ symmetry, where each $\mathbb{Z}_2$ acts by flipping the sign of one of the square roots, and this symmetry group double-covers the two solutions for $\mathbf{h}_{6,\pm}^{-1}$. Given that the Gram determinant condition is automatically satisfied on $\mathcal{L}_6$, this double covering implies that one of $\mathbf{h}_{6,\pm}^{-1}$ corresponds to  the composition $\mathbf{p}_{6}\circ\mathbf{g}_{6,1}^{-1}=\mathbf{p}_{6}\circ\mathbf{g}_{6,4}^{-1}$ while the other corresponds to $\mathbf{p}_{6}\circ\mathbf{g}_{6,2}^{-1}=\mathbf{p}_{6}\circ\mathbf{g}_{6,3}^{-1}$. It then follows that \eqref{eq:ratio.6} holds for all differential forms $\beta$ on $\mathcal{K}_6$.

\subsection{Beyond \texorpdfstring{$n=6$}{}}

We have also checked that the relations \eqref{sum.omega} and \eqref{sum.nu} are true in the $n=7$ case. In particular, we found that
\begin{align}
&\omega_7 =\omega_{7,2}+\omega_{7,3}+\omega_{7,4}+\omega_{7,5} \,, \\
&  \nu_{7,2} + \nu_{7,3} + \nu_{7,4}+ \nu_{7,5} = 2 \nu_7\,.
\label{eq:forms7}
\end{align}
The explicit forms for the differential forms present in these relations are very involved and therefore we do not provide more explicit details here. We conjecture that the relations \eqref{sum.omega} and \eqref{sum.nu} hold true for any $n$. 

Finally, we would like to add that the reduced momentum amplituhedron forms for MHV and $\overline{\text{MHV}}$ amplitudes have very simple expressions for general $n$: 
\begin{equation}
\omega_{n,2}=\bigwedge_{i=2}^{n-3} d\mathrm{log}R_{1\, i\, i+1\,i+2}\,,\qquad\qquad \omega_{n,n-2}=\bigwedge_{i=2}^{n-3} d\mathrm{log}\bar{R}_{1\, i\, i+1\,i+2}\,.
\end{equation}
These formulae are easily proven using the inverse-soft construction (see section \ref{sec:ISF}) where in both cases particle $2$ is taken to be the inverse-soft particle.


\section{Inverse-Soft Construction for Reduced Momentum Amplituhedron Forms}
\label{sec:ISF}

In this section we show how the \emph{inverse-soft (IS) construction} given in \cite{He:2018okq} can be used to obtain the reduced momentum amplituhedron form associated with any momentum amplituhedron canonical form. In particular, we present an algorithm for recursively constructing any\footnote{There exists a particular BCFW recursion scheme for which the on-shell diagrams for any tree-level amplitude are inverse-soft constructible \cite{Bourjaily:2010wh,He:2018okq,Arkani-Hamed:2016byb}.} reduced form starting from $\omega_{4,2}$ (the reduced form associated with $\Omega_{4,2}$). To this end, we first review the IS construction from \cite{He:2018okq}  and then study the effect of removing the little group scaling. 

Recall that in the BCFW construction of scattering amplitudes, the amplitude $A_{n,k}$ is given by a sum of BCFW terms and each BCFW term can be labelled by an affine permutation corresponding to a cell in the positive Grassmannian $G_{+}(k,n)$; see \cite{Arkani-Hamed:2016byb} for details.  A BCFW term for $n\ge4$ particles labelled by an affine permutation $\sigma$ is said to be \emph{inverse-soft (IS) constructible} if there exists an $i\in[n]=\{1,2,\ldots,n\}$ such that
\begin{align}
	\sigma(i-1)=i+1\,,&&
	\text{or}&&
	\sigma(i+1)=i-1\,,
\end{align}
where equality is understood to mean modulo $n$. If $\sigma(i-1)=i+1$, then $i$ is said to label a \emph{helicity-preserving IS particle}. It was argued in \cite{He:2018okq} that the corresponding canonical form can then be written as 
\begin{align}\label{IS.preserving}
	\Omega_{\sigma}(1,\ldots,i,\ldots,n)=\Omega_{\hat{\sigma}}(1,\ldots,\widehat{i-1},\widehat{i+1},\ldots,n)\wedge\Omega_{3,2}(i-1,i,i+1)\,,
\end{align}
where
\begin{align}
	\tilde{\lambda}_{\widehat{i-1}} = \tilde{\lambda}_{i-1}+\frac{\langle i\,i+1\rangle}{\langle i-1\,i+1\rangle}\tilde{\lambda}_i\,,&&
	\tilde{\lambda}_{\widehat{i+1}} = \tilde{\lambda}_{i+1}+\frac{\langle i-1\,i\rangle}{\langle i-1\,i+1\rangle}\tilde{\lambda}_i\,,
\end{align}
with $\lambda$'s unchanged and 
\begin{align}\label{is.omega-3-2}
	\Omega_{3,2}(i-1,i,i+1)=d\log\frac{\langle i-1\,i\rangle}{\langle i-1\,i+1\rangle}\wedge d\log\frac{\langle i\,i+1\rangle}{\langle i-1\,i+1\rangle}\,.
\end{align}
Alternatively, if $\sigma(i+1)=i-1$, then $i$ is a \emph{helicity-increasing IS particle} and the corresponding canonical form is given by
\begin{align}\label{IS.increasing}
	\Omega_{\sigma}(1,\ldots,i,\ldots,n)=\Omega_{\hat{\sigma}}(1,\ldots,\widehat{i-1},\widehat{i+1},\ldots,n)\wedge\Omega_{3,1}(i-1,i,i+1)\,,
\end{align}
where
\begin{align}
	{\lambda}_{\widehat{i-1}} = 	{\lambda}_{i-1}+\frac{[i\,i+1]}{[i-1\,i+1]}{\lambda}_i\,,&&
	{\lambda}_{\widehat{i+1}} = 	{\lambda}_{i+1}+\frac{[i-1\,i]}{[i-1\,i+1]}{\lambda}_i\,,
\end{align}
with $\tilde\lambda$'s unchanged and 
\begin{align}\label{is.omega-3-1}
	\Omega_{3,1}(i-1,i,i+1)=d\log\frac{[i-1\,i]}{[i-1\,i+1]}\wedge d\log\frac{[i\,i+1]}{[i-1\,i+1]}\,.
\end{align}
In both cases, $\hat{\sigma}$ is an affine permutation on $[n]\setminus\{i\}$ whose precise definition (which can be found in \cite{He:2018okq}) depends on whether $i$ is $k$-preserving or $k$-increasing. More importantly, by construction we obtain an expression for $\Omega_{\sigma}$ as a single wedge product of $d\log$'s:
\begin{align}
	\Omega_{\sigma}(1,\ldots,n)=\bigwedge_{j=1}^{2n-4}d\log\alpha_j\,.
\end{align}
We shall refer to the arguments $\{\alpha_j\}_{j=1}^{n}$ of the $d\log$'s in the above expression as \emph{canonical variables} for $\Omega_\sigma$.

Let us now consider the effect of little group scaling on canonical variables. Recall the expression for $\Omega_{4,2}$ given 
in \eqref{eq.momamp.4} and notice that
under little group scaling, the canonical variables have the following behaviour
\begin{align}
	\alpha_1 = \frac{\langle 12 \rangle}{\langle 13\rangle} \sim\frac{t_2}{t_3}\,,&&\alpha_2 = \frac{\langle 23 \rangle}{\langle 13\rangle} \sim\frac{t_2}{t_1}\,,&&\alpha_3 = \frac{\langle 34\rangle}{\langle 13\rangle} \sim\frac{t_4}{t_1}\,,&&\alpha_4= \frac{\langle 14 \rangle}{\langle 13\rangle} \sim\frac{t_4}{t_3}\,.
\end{align}
In particular, we find that for all $i\in[4]$, there is a canonical variable  $\tilde{\alpha}_{i}$ such that either $\tilde{\alpha}_{i}$ or $1/\tilde{\alpha}_{i}$ scales like $t_i/t_{i+1}$ (where $t_{4+1}=t_{1}$). This is a general property of canonical variables in IS-constructible canonical forms. In particular, if we take $\Omega_\sigma$ to be an IS-constructible canonical form on $n\ge4$ particles with canonical variables $\{\alpha_j\}_{j=1}^{2n-4}$ then for all $i\in[n]$ there is at least one canonical variable $\tilde{\alpha}_{i}\in\{\alpha_j\}_{j=1}^{2n-4}$ such that either $\tilde{\alpha}_{i}$ or $1/\tilde{\alpha}_{i}$ scales like $t_i/t_{i+1}$ (where $t_{n+1}=t_{1}$).

We can now use this fact to construct the reduced canonical forms. First, we combine the expressions \eqref{IS.preserving} and \eqref{IS.increasing} into a single formula
\begin{align}\label{IS.together}
		\Omega_{\sigma}(1,\ldots,i,\ldots,n)=\Omega_{\hat{\sigma}}(1,\ldots,\widehat{i-1},\widehat{i+1},\ldots,n)\wedge\Omega_{3,k'}(i-1,i,i+1)\,,
	\end{align}
where $k'=1$ or $k'=2$, and
\begin{align}
	\Omega_{3,k'}(i-1,i,i+1)=d\log(x_i)\wedge d\log(y_i) \,,
\end{align}
with 
\begin{align}
\begin{cases}
		x_i=\frac{[i-1\,i]}{[i-1\,i+1]}\text{ and }y_i=\frac{[i\,i+1]}{[i-1\,i+1]},& \text{for }k'=1\,,\\
		x_i=\frac{\langle i-1\,i\rangle}{\langle i-1\,i+1\rangle}\text{ and }y_i=\frac{\langle i\,i+1\rangle}{\langle i-1\,i+1\rangle},& \text{for }k'=2\,.
	\end{cases}
\end{align}
Let us denote by $\{\beta_j\}_{j=1}^{2n-6}$ the canonical variables for $\Omega_{\hat{\sigma}}$ in \eqref{IS.together}. Then there exists a $\tilde{\beta}\in\{\beta_j\}_{j=1}^{2n-6}$ such that
	\begin{align}
		\omega_\sigma(1,\ldots,i,\ldots,n)=(-1)^{n+i+k'}d\log\left(\frac{x_i}{y_i}\tilde{\beta}^{s}\right)\wedge\omega_{\hat{\sigma}}(1,\ldots,\widehat{i-1},\widehat{i+1},\ldots,n)\,.
	\end{align}
Here, $\tilde\beta$ and $s\in\{\pm 1\}$ are fixed by the requirement that the argument of the first logarithm, $\frac{x_i}{y_i}\tilde{\beta}^{s}$, is little group scaling invariant. In particular, since $\frac{x_i}{y_i}\sim \frac{t_{i+1}}{t_{i-1}}$ for $k'=1$ and $\frac{x_i}{y_i}\sim \frac{t_{i-1}}{t_{i+1}}$ for $k'=2$, we know from our discussion above that we can always find such a $\tilde\beta$ which cancels this scaling \footnote{Notice that $\Omega_{\hat\sigma}$ does not depend on particle $i$ which means that there is always at least one canonical variable which will scale as $\frac{t_{i-1}}{t_{i+1}}$ or $\frac{t_{i+1}}{t_{i-1}}$.}. 

We can fix all MHV reduced forms to have coefficient $+1$ by choosing $i=2$ to be the IS particle and by absorbing $(-1)^n$ into a redefinition of $\Omega_n$. This is the origin of the minus signs in \eqref{eq.momamp.5} and \eqref{eq.momamp.6}. 
Fixing the sign of the MHV reduced form in this way provides a useful prescription for fixing the sign ambiguity in the definition of the canonical form of the kinematic associahedron.

\section{Conclusions}

Scattering amplitudes in various theories are encoded in logarithmic differential forms on the kinematic space.   
In this paper we have shown a surprising relation between the canonical forms of the momentum amplituhedron and the kinematic associahedron, i.e.~the positive geometries associated to tree-level amplitudes in $\mathcal{N}=4$ sYM and bi-adjoint $\phi^3$ theory, respectively.   In particular, starting from the differential form for the full amplitude in  $\mathcal{N}=4$ sYM and stripping off the (highest-degree) little group scaling dependence we find the associahedron form. This relation exposes the singularities of the respective scattering amplitudes and captures the fact that the factorization channels, corresponding to vanishing planar Mandelstam variables, are the same. 

The relation we found is at the level of differential forms. The most natural and interesting question is how to relate the geometries themselves: whether there exists a map which directly connects boundaries of the momentum amplituhedron to boundaries of the kinematic associahedron. While we understand the relation between boundaries corresponding to multiparticle poles,  we lack a systematic understanding of the two-particle pole boundaries.
Indeed, collinear singularities $s_{i,i+1}\rightarrow 0$ correspond to one boundary of the associahedron, while there are two boundaries of the momentum amplituhedron associated to them, since $s_{i,i+1} = \langle i i+1\rangle [i i+1] \rightarrow 0$ can be reached by setting either $\langle i i+1\rangle \rightarrow 0$ or $ [i i+1] \rightarrow 0$, see \cite{Ferro:2020lgp} for details.
Related to this, it would be interesting to understand if and how we can interpret the pull-back of the differential form associated to the full superamplitude $\Omega_n$ as the ``product" of two geometries: one associated to the projective space $\mathbb{P}^{n-1}$ and one associated to the pull-back of the associahedron form $\tilde\omega_n$. Finally, we do not know how the sum over different helicity sectors of the reduced momentum amplituhedron form combine to describe the pull-back of the associahedron form  from a geometric point of view. In particular, we do not know whether it is a triangulation  or a superposition of geometries. 
We leave these important points to future work.

\section{Acknowledgements}

We would like to thank Nima Arkani-Hamed for useful discussions.
This work was partially funded by the Deutsche Forschungsgemeinschaft (DFG, German Research Foundation) -- Projektnummern 404358295 and 404362017.

\appendix
\section{Extended Fock-Goncharov Parametrization for Five-point Amplitudes{}}\label{app:tildelambda}

The $\tilde\lambda$ matrix in the extended Fock-Goncharov parametrization reads
\begin{equation}\label{tildelambda5}
\tilde\lambda=\left(
\begin{array}{ccccc}
 t_1^{-1} & 0 &  t_3^{-1}\tilde\lambda_3^1 & t_4^{-1} \tilde\lambda_4^1&  
 t_5^{-1}\tilde\lambda_5^1   \\
 0 &  t_2^{-1} s_{1,2} & t_3^{-1}\tilde\lambda_3^2 &  t_4^{-1}\tilde\lambda_4^2&
 t_5^{-1}\tilde\lambda_5^2 \\
\end{array}
\right)
\end{equation}
with
\begin{align}
\tilde\lambda_3^1&=\frac{R_{1234} (R_{1345} (\bar{R}_{1234}+1)-\bar{R}_{1234} \bar{R}_{1345})-\bar{R}_{1234}
   \bar{R}_{1345}}{R_{1234} \bar{R}_{1234} ((R_{1234}+1) \bar{R}_{1345}-R_{1345} (-R_{1234}
   \bar{R}_{1345}+\bar{R}_{1234} (\bar{R}_{1345}+1)+1))}\\
 \tilde\lambda_4^1&=  \frac{(\bar{R}_{1234}+1) (\bar{R}_{1234}
   (\bar{R}_{1345}+1)-R_{1234} (R_{1345}+1))}{R_{1234} \bar{R}_{1234} ((R_{1234}+1)
   \bar{R}_{1345}-R_{1345} (-R_{1234} \bar{R}_{1345}+\bar{R}_{1234} (\bar{R}_{1345}+1)+1))} \\
 \tilde\lambda_5^1&= \frac{(\bar{R}_{1234} (\bar{R}_{1345}+1)+1) (R_{1234}-\bar{R}_{1234})}{R_{1234} \bar{R}_{1234}
   ((R_{1234}+1) \bar{R}_{1345}-R_{1345} (-R_{1234} \bar{R}_{1345}+\bar{R}_{1234} (\bar{R}_{1345}+1)+1))}\\
\tilde\lambda_3^2&=   -\frac{s_{1,2} (R_{1234} (R_{1345} (\bar{R}_{1234}+1)-\bar{R}_{1234}
   \bar{R}_{1345})-\bar{R}_{1234} \bar{R}_{1345})}{R_{1234} \bar{R}_{1234} (R_{1345}-\bar{R}_{1345})}\\
\tilde\lambda_4^2&=    \frac{s_{1,2} (R_{1234} (R_{1345}+1)-\bar{R}_{1234} (\bar{R}_{1345}+1))}{R_{1234} \bar{R}_{1234}
   (R_{1345}-\bar{R}_{1345})} \\
 \tilde\lambda_5^2&=  \frac{s_{1,2} (\bar{R}_{1234}-R_{1234})}{R_{1234} \bar{R}_{1234}
   (R_{1345}-\bar{R}_{1345})} \,.
\end{align}
\section{Formulae for Six-point Amplitudes}
\label{app:n6}

The arguments of the square roots, $\Delta_1$ and $\Delta_2$, appearing in the four solutions for the inverse of the map $\mathbf{g}_6$, see \eqref{p6inverse}, read explicitly:
\begin{align}\nonumber
\Delta_1&=s_{1,2}^2 (s_{2,3}-s_{2,3,4})^2+(s_{2,3} (s_{3,4}-s_{5,6})+s_{1,2,3} (s_{2,3,4}-s_{3,4}))^2+\\\nonumber
&+2 s_{1,2} \left(-(s_{3,4}+s_{5,6})
   s_{2,3}^2+(s_{3,4} (s_{1,2,3}-2 s_{5,6})+(s_{3,4}+s_{5,6}+s_{1,2,3}) s_{2,3,4}) s_{2,3}+\right.\\
   &+\left. s_{1,2,3} (s_{3,4}-s_{2,3,4}) s_{2,3,4}\right)\label{Delta1}\,,\\ \nonumber
   \Delta_2&=(s_{2,3} s_{5,6}-s_{6,1} s_{5,6}+s_{6,1} s_{1,2,3}-s_{1,2,3} s_{2,3,4}+s_{4,5} (s_{2,3,4}-s_{5,6}))^2-\\&-4 s_{4,5} s_{5,6} s_{6,1}
   (s_{2,3}+s_{5,6}-s_{1,2,3}-s_{2,3,4}) \,,
   \label{Delta2}
\end{align}
while the explicit form for $\Gamma$ appearing in the solutions for the inverse of the map $\mathbf{h}_6$ in \eqref{Gram6} is
\begin{align}
	\nonumber \Gamma&=\left(s_{3,4}-s_{5,6}\right) s_{5,6} s_{2,3}^2+\Big(s_{5,6} \left(s_{5,6} s_{6,1}+s_{4,5} \left(s_{5,6}-s_{2,3,4}\right)-s_{1,2,3} \left(s_{6,1}-2 s_{2,3,4}\right)\right)\\
	\nonumber &-s_{3,4} \left(s_{5,6} \left(s_{6,1}+s_{1,2,3}\right)+s_{4,5} \left(s_{5,6}-s_{2,3,4}\right)+s_{1,2,3} \left(s_{6,1}+s_{2,3,4}\right)\right)\Big) s_{2,3}\\
	\nonumber &+s_{1,2,3} \Big(\left(s_{4,5}-s_{1,2,3}\right) s_{2,3,4}^2+\big(-s_{5,6} \left(s_{4,5}+s_{6,1}\right)+s_{6,1} s_{1,2,3}\\
	\nonumber &+s_{3,4} \left(-s_{4,5}+2 s_{6,1}+s_{1,2,3}\right)\big) s_{2,3,4}+s_{3,4} \left(s_{4,5} s_{5,6}+s_{6,1} \left(s_{1,2,3}-s_{5,6}\right)\right)\Big)\\
	\nonumber &+s_{1,2} \Big(\left(s_{4,5}+s_{1,2,3}\right) s_{2,3,4}^2-\big(s_{5,6} \left(s_{4,5}-s_{6,1}\right)+\left(s_{6,1}-2 s_{4,5}\right) s_{1,2,3}\\
	&+s_{2,3} \left(s_{4,5}+s_{5,6}+s_{1,2,3}\right)\big) s_{2,3,4}+s_{2,3} \left(s_{5,6} \left(s_{2,3}-s_{4,5}-s_{6,1}\right)+s_{6,1} s_{1,2,3}\right)\Big).
\end{align}

\section{Geometry of the Differential Form \texorpdfstring{$\nu_{6,2}$}{the reduced momentum amplituhedron form for (n,k)=(6,2) on the Gram determinant surface}}
\label{app:nu62}
In performing the push-forward $(\mathbf{g}_6)_\ast\,\omega_{6,2}=\nu_{6,2}$, the square-roots present in the individual solutions $\mathbf{g}_{6,i}^{-1}$ disappear in the sum over all four solutions, and we obtain 
\begin{align}
\nonumber \nu_{6,2}=&\nu_{6,2}^{(A)}(s_{1,2},s_{2,3},s_{3,4};s_{4,5},s_{5,6},s_{6,1})+\nu_{6,2}^{(B)}(s_{1,2},s_{4,5},s_{1,2,3};s_{3,4},s_{6,1},s_{2,3,4})\\
-&\nu_{6,2}^{(A)}(s_{4,5},s_{5,6},s_{6,1};s_{1,2},s_{2,3},s_{3,4})-\nu_{6,2}^{(B)}(s_{3,4},s_{6,1},s_{2,3,4};s_{1,2},s_{4,5},s_{1,2,3}),
\label{nu-6-2}
\end{align}
where
\begin{align}
	\nonumber \nu_{6,2}^{(A)}&(s_{1,2},s_{2,3},s_{3,4};s_{4,5},s_{5,6},s_{6,1})=d\log s_{1,2}\wedge d\log s_{3,4}\wedge d\log\left(\frac{s_{4,5} s_{6,1}}{s_{1,4} s_{5,6}}\right)\\
	 &+\frac{s_{2,3}}{s_{1,2,3}-s_{2,3,4}} d\log\left(\frac{s_{1,2} s_{3,4}}{s_{2,3}^2}\right)\wedge d\log\left(\frac{s_{1,4}}{s_{2,3}}\right)\wedge d\log\left(\frac{Q}{s_{1,4} s_{5,6}}\right)\,,\\
	 \nonumber \nu_{6,2}^{(B)}&(s_{1,2},s_{4,5},s_{1,2,3};s_{3,4},s_{6,1},s_{2,3,4})=\frac{s_{1,2,3}}{s_{1,2,3}-s_{2,3,4}}\\\nonumber &\times\Bigg\{\Bigg[d\log\left(\frac{s_{4,5}}{s_{1,2}}\right)\wedge d\log\left(s_{1,4} s_{2,3}\right)-d\log\left(\frac{s_{1,2}}{s_{5,6}}\right)\wedge d\log\left(\frac{s_{5,6}}{s_{2,3}}\right)\Bigg]\wedge d\log Q\\
	\nonumber &\phantom{\Bigg\{}+\Bigg[d\log s_{5,6}\wedge d\log\left(\frac{s_{1,2} s_{3,4}}{s_{2,3}}\right)-d\log s_{2,3}\wedge d\log\left(\frac{s_{4,5} s_{6,1}}{s_{5,6}}\right)\Bigg]\wedge d\log\left(\frac{s_{1,4}}{Q}\right)\Bigg\} \\
	&+d\log s_{1,2}\wedge d\log s_{4,5}\wedge d\log Q\,,
\end{align}
$s_{1,4}=s_{2,3}+s_{5,6}-s_{1,2,3}-s_{2,3,4}$ and $Q=s_{2,3} s_{5,6}-s_{1,2,3} s_{2,3,4}$.

It is easy to verify that $\nu_{6,2}$ has simple poles on all planar two-particle Mandelstam variables (as expected) as well as on $s_{1,4}$ and $Q$. These additional, unexpected poles are a consequence of the Gram determinant condition for six particles, and are precisely the denominator factors in the two solutions to the Gram determinant equations given in \eqref{Gram6}.

Taking subsequent residues of $\nu_{6,2}$ produces not only $\pm1$, but  also $\pm2$ on zero-dimensional boundaries, in contradistinction to canonical forms which, by definition, have residues $\pm 1$ on boundaries of zero dimension \cite{Arkani-Hamed:2017tmz}. One might have expected $\nu_{6,2}$ to be a canonical form since it is the push-forward of $\omega_{6,2}$ (a canonical form). However, the push-forward only preserves canonical forms which are top-dimensional \cite{Arkani-Hamed:2017tmz}, and since $\omega_{6,2}$, defined on $\mathcal{L}_6$, is not a top-form, we anticipate that this is the origin of the non-canonical behaviour of $\nu_{6,2}$. We have also verified that for each individual solution $\mathbf{g}_{6,i}^{-1}$, $\nu_{6,2}^{(i)}=(\mathbf{g}_{6,i}^{-1})^\ast \omega_{6,2}$ is a positive geometry with residues $\pm 1$. This suggests the simple interpretation for the geometry of $\nu_{6,2}$ as the geometric sum of the four positive geometries $\nu_{6,2}^{(i)}$.

\bibliographystyle{nb}

\bibliography{momampl_assoc}

\end{document}